\documentstyle[12pt]{article}

\def\lam{\lambda}

\def\del{\partial}

\def\dis{\displaystyle}

\def\a{\rm a}
\def\b{\rm b}

\begin{document}

\begin{center}
{\large\bf Correspondence between the first and the second order\\[2mm]
formalism in the generalized gravity of $f(R)$-type}\\[8mm]
\end{center}
\hspace*{15mm}\begin{minipage}{13.5cm}
Y. Ezawa, H. Iwasaki, Y. Ohkuwa$^{\dagger}$, N. Yamada and T. Yano$^{*}$\\[3mm]
Department of Physics, Ehime University, Matsuyama, 790-8577, Japan\\[1mm]
\hspace*{-1.5mm}$^{\dagger}$Department of Mathematics, Miyazaki Medical College, 
Kitazoe, Miyazaki\\[.5mm]
 889-1692, Japan\\[1mm]
\hspace*{-1.5mm}$^{*}$Department of Electrical Engineering, Ehime University, 
Matsuyama, 790-\\[.5mm]
8577, Japan\\[5mm]
{\small Email :  ezawa@sci.ehime-u.ac.jp, hirofumi@phys.sci.ehime-u.ac.jp,\\
ohkuwa@post.miyazaki-med.ac.jp, naohito@phys.sci.ehime-u.ac.jp\\
 and yanota@eng.ehime-u.ac.jp}
\\[8mm]
{\bf Abstract}\\[2mm]
 We clarify the correspondence between the first order formalism and the second order
formalism in the generalized gravity where the Lagrangian density is given by 
a function of the scalar curvature $f(R)$.\\

\end{minipage}

\noindent
{\it Introduction}:
Up to the present, our knowledge of the universe has become wider both 
observationally and theoretically.
So the theories of gravity, the dominant force in the universe, have been extended
and variety of generalized theories of gravity have been proposed, starting from 
the Einstein one where the Lagrangian density is a linear function of the scalar 
curvature.
The natural motivation arises when we consider the initial singularity of our 
universe\cite{EH,Wald} where the gravity should be so strong that higher powers of 
the invariants constructed from the scalar curvature, the Ricci tensor and 
the Riemann tensor, would play important roles\cite{N-NT-TAN}.
Examples are given when we consider the quantum field theory in a curved spacetime
\cite{BD}, the effects of gravity on strings, extended objects, since they receive 
the tidal force described by the curvature.
The local expression of the latter leads to the modification of gravity by higher 
powers of the curvature tensors.
The string perturbation theory, in fact, gives a definite combination known as 
the Gauss-Bonnet combination\cite{Zwie}.

In this work we confine ourselves to a type of theories whose Lagrangian density is 
a function of the scalar curvature $f(R)$.
This type of theory has long been investigated and is expected to resolve 
the initial singularity problem\cite{N-NT-TAN,S-S,BHV,EIOUYY}.
Recently a theory including the inverse power of the scalar curvature was proposed 
and discussed\cite{CCTetc,VMF} to explain the present accelerating universe
\cite{SNIa,CMB}.
Thus this type of theory is a convenient generalization of the Einstein one both to 
strong  and weak gravity region.

Theoretically conformal equivalence of this theory to the Einstein one are well 
known\cite{MS,CMQ}.
However, identification of the physical metric has not been settled.
Other theoretical problem seems to be a more fundamental one, i.e. how to apply 
the variational principle.
It is known that the first order, or Palatini, formalism of Einstein gravity is 
equivalent to the second order formalism.
We mean by the second order formalism the one where the connection is given by 
the Christoffel symbol, so that the metric tensor is the only variable describing 
gravity.
In this sense, the first order formalism is another line of generalizing Einstein 
gravity, since the connection is not given {\it a priori}, but is determined by 
the physical law, the variational principle.
The above equivalence means that the relation of the connection to the Christoffel 
symbol determined by the variational principle is that they are the same.
However, for general $f(R)$-type gravity, they are different.
This was noticed from the fact that the connection is not the Christoffel symbol 
but is given by e.g. the relation (3) below\cite{S-S,BHV,VMF,CMQ}.
Recently Flanagan showed that this difference is also seen by transforming to 
the conformally related Einstein frame where the scalar field has no kinematical 
term contrary to the  second order formalism\cite{VMF}.
Here we will present another way of expressing the difference by clarifying 
the correspondence of the first order formalism and the second order formalism by 
constructing the Lagrangian density to be used in the second order formalism 
starting from the first order Lagrangian density.
In doing so we use the fact that the connection in the first order formalism is just
the conformal transformation of the Christoffel symbol as seen in (3a).\\
{\it Correspondence}:
We consider a theory described by an action
$$
S=\int \sqrt{-g}f(R)d^4x                                           \eqno(1)
$$
where
$$
R=g^{\mu\nu}R_{\mu\nu},\ \ \ g=\det g_{\mu\nu}.                    \eqno(2)
$$
$g^{\mu\nu}$ is the metric tensor and $R_{\mu\nu}$ the Ricci tensor.
In the first order formalism $R_{\mu\nu}$ is given by the usual expression 
in terms of the connection $\Gamma^{\lam}_{\mu\nu}$ taken as independent variables.
This formalism is in line with the generalization of gravity since the metric and the connection are independent concepts geometrically.
The variation with respect to $\Gamma^{\lam}_{\mu\nu}$ leads to the relation
$$
\Gamma^{\lam}_{\mu\nu}
=\Gamma^{\lam(0)}_{\mu\nu}
+{1\over 2f'(R)}\left[\delta^{\lam}_{\mu}\del_{\nu}f'(R)
+\delta^{\lam}_{\nu}\del_{\mu}f'(R)-g_{\mu\nu}\del^{\lam}f'(R)\right]
                                                                   \eqno(3\a)
$$
or
$$
\Gamma^{\lam}_{\mu\nu}
=\Gamma^{\lam(0)}_{\mu\nu}
+{f''(R)\over 2f'(R)}\left[\delta^{\lam}_{\mu}\del_{\nu}R
+\delta^{\lam}_{\nu}\del_{\mu}R-g_{\mu\nu}\del^{\lam}R\right]
                                                                   \eqno(3\b)
$$
where superscript $^{(0)}$ indicates that the quantity is constructed from 
the metric, i.e. the quantity to be used in the second order formalism or 
metric compatible quantities.
Here $\Gamma^{\lam(0)}_{\mu\nu}$ is the Christoffel symbol constructed from 
the metric $g_{\mu\nu}$.
It is clear that the connection $\Gamma^{\lam}_{\mu\nu}$ is different from 
the Christoffel symbol $\Gamma^{\lam(0)}_{\mu\nu}$ except for Einstein gravity 
where $f''(R)=0$ and the case where $R=constant$, i.e. $\del_{\mu}R=0$.

Now we consider a conformal transformation
$$
\bar{g}_{\mu\nu}=\Omega^2g_{\mu\nu},\ \ \ 
\Omega^2=k\times f'(R),\ (k=constant).                             \eqno(4)
$$
The constant $k$ is chosen such that $\Omega=1$ for Einstein gravity. 
Then the relation (3a) is known to be the conformal transformation of 
the Christoffel symbol:
$$
\Gamma^{\lam}_{\mu\nu}=\bar{\Gamma}^{\lam(0)}_{\mu\nu}             \eqno(5)
$$
This means that the theory becomes the second order if $\bar{g}_{\mu\nu}$ is used 
as the metric tensor.
The transition is straightforward as is shown below.
Thus introduction of the connection as independent variables to describe gravity 
seems to be redundant.
The action in the second order formalism is taken to be the action (1) rewritten in 
terms of the metric $\bar{g}_{\mu\nu}$.
Equation (5) means 
$$
R_{\mu\nu}=\bar{R}_{\mu\nu}^{(0)}
$$
which in turn means
$$
R=kf'(R)\bar{R}^{(0)}.                                             \eqno(6)
$$
Note that the contraction is taken between the untransformed metric and 
the transformed Ricci tensor.
Solving the equation (6) with respect to $R$, we can write
$$
R=\xi(\bar{R}^{(0)}).                                              \eqno(7)
$$
In conformally transformed frame, only metric compatible quantities will be used, 
so we drop the superscript $^{(0)}$ in this frame hereafter.
Noting that $g=\bar{g}/k^4f^{'4}(R)=\bar{g}/k^4f^{'4}(\xi(\bar{R}))$, we can rewrite
 (1) as
$$
S=\int\sqrt{-\bar{g}}F(\bar{R})                                   \eqno(8)
$$
where
$$
F(\bar{R})
={f\left(\xi(\bar{R})\right)\over k^2f^{'2}\left(\xi(\bar{R})\right)}\eqno(9\a)
$$
which is also expressed, in a more symmetrical form, as
$$
{F(\bar{R})\over \bar{R}^2}={f(R)\over R^2}                        \eqno(9\b)
$$
Equation (8) gives the action to be used in the second order formalism corresponding
 to the action (1) to be used in the first order formalism.\\
The functional form of $f(R)$ and $F(\bar{R})$ are different in general.
This is another way of indicating that the second order formalism, in which 
the action (1) is taken as $f(R^{(0)})$, would be different from the one in 
the first order formalism.
We will refer the frame in which the metric is $\bar{g}_{\mu\nu}$ as the second 
order frame.

Finally we will illustrate the correspondence by examples for $f(R)$:\\
(i) $f(R)=a+bR+cR^2$

Equation (6) is a linear equation for $R$ which is solved to give
$$
R=\bar{R}/(1-{2c\over b}\bar{R})
$$
where $k$ is chosen to satisfy $kb=1$.
Then from (9), we obtain
$$
F(\bar{R})=a+\left(1-{4ac\over b^2}\right)(b\bar{R}-c\bar{R}^2).   \eqno(10)
$$
(ii) $f(R)=a+bR+cR^2+dR^3$

Equation (6) is a quadratic equation for $R$ which is solved to give
$$
R={1\over 6d\bar{R}}
\left[b-2c\bar{R}\pm \sqrt{(b-2c\bar{R})^2-12bd\bar{R}^2}\;\right]
$$
where $k$ is again chosen to satisfy $kb=1$.
$F(\bar{R})$ is expressed as
$$
\begin{array}{lcl}
F(\bar{R})&=&\dis {1\over2}a+{2\over3}b\left(1-{3ac\over b^2}\right)\bar{R}
           -{c\over3}\left(1-{6ac\over b^2}+{9ad\over bc}\right)\bar{R}^2\\[5mm]
&\pm&\dis \left[{1\over2}a+{1\over3}b\left(1-{3ac\over b^2}\right)\bar{R}\right]
\sqrt{\left(1-{2c\over b}\bar{R}\right)^2-{12d\over b}\bar{R}^2}
\end{array}                                                        \eqno(11)
$$
(iii) $f(R)=a+bR+c/R$

In this case, equation (6) becomes a cubic equation
$$
R^3-kb\bar{R}R^2+kc\bar{R}=0.                                      \eqno(12)
$$
The solutions of this equation are known to be given by a set of 
formulae which are not cited here.
We only note that (12) has a real solutions if $c/b$ is negative or
$|\bar{R}|^2<4c/27b$ 3 real solutions otherwise.
Thus for this type of $f(R)$, although the second order form could also be 
described by $f(R^{(0)})$, the results differ qualitatively from the case when 
$f(R)$ is used in the first order formalism.\\
{\it Concluding remarks}: We clarified the correspondence between the first order 
formalism and the second order formalism, i.e. for the Lagrangian density 
$\sqrt{-g}f(R)$ of the first order formalism, we identified the Lagrangian density 
$\sqrt{-\bar{g}}F(\bar{R})$ to be used in the second order formalism.
The functional form of $f(R)$ and $F(\bar{R})$ are different except for 
the Einstein case.
It is noted, however, for quadratic $f(R)$, $F(\bar{R})$ is also quadratic although 
the coefficients are different.

The Lagrangian density $\sqrt{-\bar{g}}F(\bar{R})$ may in turn be conformally 
transformed, so that it describe Einstein gravity with a scalar field.
We would like to comment on this conformal equivalence.
If we take the relevant conformal transformation to be a change of variables, 
equivalence is clear except that which metric should be identified as, e.g. FRW 
metric.
However, when the dynamical aspects are concerned, situation is not so clear 
because the transformation depends on the curvature.
The curvature includes the second order derivative of the metric.
From the viewpoint of canonical formalism, the transformation mixes the coordinates 
with the momenta, i.e. the transformation is the one in the phase space.
Therefore, unless the transformation turns out to be a canonical one, the dynamical 
system is changed by the transformation.
The canonical transformation of the higher-curvature gravity has some subtle points.
Concerning the physical nature of the metric, a new problem arises whether 
the metric $\bar{g}_{\mu\nu}$ in the second order frame is physical or not.
These problems will be investigated in separate works.

\end{document}